# On the application of Good-Turing statistics to quantify convergence of biomolecular simulations.

Panagiotis I. Koukos & Nicholas M. Glykos*

*Department of Molecular Biology and Genetics, Democritus University of Thrace, University campus, 68100 Alexandroupolis, Greece, Tel +30-25510-30620, Fax +30-25510-30620, http://utopia.duth.gr/~glykos/ , glykos@mbg.duth.gr*

# Abstract

Quantifying convergence and sufficient sampling of macromolecular molecular dynamics simulations is more often than not a source of controversy (and of various *ad hoc* solutions) in the field. Clearly, the only reasonable, consistent and satisfying way to infer convergence (or otherwise) of a molecular dynamics trajectory must be based on probability theory. Ideally, the question we would wish to answer is the following : "What is the probability that a molecular configuration important for the analysis in hand has not yet been observed ?". Here we propose a method for answering a variant of this question by using the Good-Turing formalism for frequency estimation of unobserved species in a sample. Although several approaches may be followed in order to deal with the problem of discretizing the configurational space, for this work we use the classical RMSD matrix as a means to answering the following question : "What is the probability that a molecular configuration with an RMSD (from all other already observed configurations) higher than a given threshold has not actually been observed ?". We apply the proposed method to several different trajectories and show that the procedure appears to be both computationally stable and internally consistent. A free, open-source program implementing these ideas is immediately available for download *via* public repositories.

# Running title

Quantifying convergence of MD trajectories

# 1. Introduction

Even a cursory examination of recent molecular dynamics literature shows that the treatment of convergence (or sufficient sampling) of the corresponding simulations can follow either of two distinct paths. The first is to ignore the subject altogether in the hope that the quoted simulation times will appear to be so long that no further evidence of convergence will be required. The second is to select one (or more) of the various methods currently available such as eigenspace overlap or cosine content (see Grossfield and Zuckerman[1] for an excellent review) and apply them in the hope that the outcome (for example, an eigenspace overlap of 0.85 between two independent halves of the trajectory using the top three principal components) will appear to be so overwhelmingly convincing that no further quantification of convergence will be necessary. Although several of these methods can and do serve their purpose, i.e. they can meaningfully quantify the extent of sampling of the corresponding trajectories, we do feel that a proper probabilistic measure of convergence is the only consistent and satisfying method of inference for something as inherently probabilistic as is the (necessarily limited) sampling of a molecular dynamics trajectory.

However, stating that a probabilistic treatment is the only natural solution to a problem, does little to aid its solution. Naturally, the first question that must be answered is how to define the problem of convergence in probabilistic terms. We are convinced that the only reasonable answer to this question must be pragmatic, i.e. directly related to the sought analysis of the simulations' data. We believe that what we should like to calculate is the probability that a molecular configuration important for a given analysis has not yet been observed. Defining, however, what is "important for the analysis in hand" is more difficult. For this work we have decided to quantify structural distance (and, thus, importance) using the possibly most popular measure of structural similarity, the root-mean-square deviation (RMSD). This choice greatly simplifies both the treatment of the problem and its implementation as follows. In the first step an RMSD matrix is constructed from the trajectory, possibly using only selected substructures depending on the analysis performed (if, for example, the aim of the analysis concerns only the dynamics of an enzyme's active site residues, then only these residues would be used for constructing the matrix). In the second step the RMSD matrix is being treated as a distance matrix and a dendrogram is constructed using established hierarchical clustering methods[2]. In the third step the dendrogram is used to produce clusters at various RMSD cutoffs together with their frequencies. In the final step, the Good-Turing formalism[3,4] is applied to these sets of frequencies to obtain an estimate of the probability of unobserved species. The result of the proposed method is a table of the form "RMSD threshold vs. Probability unobserved" which allows a direct and immediate representation of the answer to the question "what is the probability that structures with RMSDs higher than a given threshold have not yet been observed in the simulation ?". The expression "structures with RMSDs higher than a given threshold" will be used throughout this communication, and to avoid confusion we should define explicitly which 'RMSD' we refer to : The quoted RMSD should be understood to be the smallest of all RMSDs between a new (previously unobserved) structure and the set of structures already observed in the trajectory that is being analyzed. To make this even more definite : if for an RMSD threshold of 1.0 Å the probability of unobserved species is, say, 0.40, then this should be understood to mean that if we were to continue the simulation, then we would expect 40% of the new (previously unobserved) structures to differ by *at least* 1.0 Å (RMSD) from all already observed trajectory structures (or, equivalently, that 60% are expected to differ by less than 1.0 Å RMSD).

In the following paragraphs we describe in more detail the principal ideas behind the method and the actual algorithm encoded in the computer program that we distribute. This is followed by an extensive discussion of results obtained from the application of the program to several different biomolecular trajectories. We close by discussing the practical limitations arising from the usage of RMSD matrices, and other possible approaches to the problem.

## 2. Methods, algorithms and implementation

### 2.1 Outline of the method and algorithms

The essence of our method is the following. We treat the molecular dynamics trajectory as a finite sample of "molecular species" (clusters of similar structures) taken from an underlying distribution containing an unknown number of such molecular species. The observed frequencies of molecular species in the sample are calculated, and the Good-Turing formalism is applied to these frequencies allowing us to estimate the total probability of unseen (i.e. as yet unobserved) species.

The implementation of this method through the application of RMSD matrices appears to be straightforward : given a molecular dynamics trajectory, construct the frame-to-frame RMSD matrix, treat the RMSD matrix as a distance matrix to construct a dendrogram, use the dendrogram to obtain frequencies of clusters for various RMSD cutoffs, treat these frequencies as frequencies of "observed molecular species" for the given RMSD cutoff, and in the final step, employ the Good-Turing formalism to estimate the probability of unobserved species for the given RMSD cutoff. The result would be the sought distribution of the (probability of unobserved species vs. RMSD). However, closer examination of the procedure described above shows that such a simple-minded application of the algorithm is bound to fail : The application of Good-Turing statistics assumes that the structures used for constructing the RMSD matrix are sufficiently distant in time, so distant that they can be treated as independent 'objects' of a sample. In other words, it is assumed that successive entries in the matrix (and the corresponding structures) are not mechanistically correlated due to the very short time interval used for recording structures from the trajectory. The implication is that a direct application of the algorithm as described above would lead to results that are dependent on how fine is the (otherwise arbitrary) sampling of the trajectory. This is clearly both highly unsatisfactory and erroneous. The important addition to the algorithm, then, is to note that it is possible –using the RMSD matrix alone and no other source of information– to correct for this dependency on the sampling interval of the original molecular dynamics trajectory using a procedure similar to the one described by Flyvbjerg and Petersen[5]. Because this is an important aspect of the method, a detailed description of this correction follows.

The crucial observation is that if the sampling of the trajectory is such that successive structures are not time-correlated, then the distribution of the maximum of the RMSDs observed between any two successive (in the matrix) structures should be independent of the sampling, and would, thus, be approximately the same even if instead of using the original NxN matrix, we used an [(N/2)x(N/2)] matrix (obtained by taking every second row and column of the original matrix, we will refer to this as a 'sampling factor of 2'), or an [(N/3)x(N/3)] matrix (obtained by taking every third row and column of the original matrix, sampling factor of 3), or [(N/4)x(N/4)] etc. Note that the application of the maximum (and, not for example, the average) of the RMSDs observed between successive structures is also intuitively correct : any two successive structures can be very much alike simply because the given molecular conformation is stable. It is the maximum of the observed RMSDs between successive frames that can tell us whether two structures are similar because they are stable, or whether all neighboring structures are similar because there was not enough simulation time for them to differentiate. With so much of an introduction, the actual numerical application of this criterion is straightforward :

1. Take the original (NxN) matrix and find the maximum of all RMSDs present in the matrix's superdiagonal (the superdiagonal is the first diagonal immediately adjacent to the principal diagonal. Because the RMSD matrices are symmetric, this is identical with the subdiagonal). This maximum RMSD is the highest RMSD observed between successive structures in the original (NxN) sampling of the trajectory (sampling factor of 1).
2. Construct a sub-matrix of size [(N/2)x(N/2)] from the original matrix by taking every second row and column. Determine the maximum of all RMSDs present on the superdiagonal of this new matrix. This maximum RMSD is the highest RMSD observed between successive structures with a sampling factor of 2 (with respect to the original). Because there are two choices of origin for constructing the [(N/2)x(N/2)] matrix, we can calculate an average value of this maximum RMSD plus its estimated standard deviation.
3. Repeat the above to obtain maximal RMSDs (plus their estimated standard deviations) for increasing values of the sampling factor.

Fig.1a illustrates the application and results obtained from this procedure. The upper (black) curve in this panel shows the distribution of (maximum RMSDs vs. sampling factors) for the whole of a 4.4 μs-long folding simulation of a 40-residue three-helix bundle. The dimensions of the corresponding RMSD matrix were 9220×9220 data points (note that for all calculations reported in this communication we have only used the proteins' $C_\alpha$ atoms for constructing the corresponding RMSD matrices). Clearly, the distribution of [max(RMSD) vs. sampling] converges very quickly to a stable maximum RMSD of approximately 20Å with a corresponding sampling factor upon convergence of ~2. The lower (red) curve shows the same distribution from the same trajectory, but using only the first 0.8 μs of the simulation and with a matrix of comparable size (10,000×10,000). The result is that the molecular dynamics trajectory corresponding to the lower (red) curve has been sampled so much more finely (than the one corresponding to the upper curve), that successive structures are highly time-correlated. The resulting graph (red curve in Fig.1a) brings this forward and clearly indicates that for small values of the sampling factor the corresponding structures are not independent, and thus can not be used in Good-Turing statistics, with convergence being reached only much later, for values of the sampling factor of ~30.

Having obtained the distribution of (maximum RMSDs vs. sampling factors) as shown in Fig.1a, the question arises how to accurately determine for which value of the sampling factor convergence of the maximal RMSDs is reached (corresponding to the plateau of the graphs). Once this value is known, the method is essentially complete : if, for example, we could determine that the optimal sampling factor is, say, 4, then we would construct the four sub-matrices of dimensions [(N/4)x(N/4)], and for each sub-matrix calculate the corresponding dendrogram, determine the (probability of unobserved species vs. RMSD), and in the final step, calculate the averages (of the four repetitions) plus their estimated standard deviations. We have chosen to tackle the problem of determining the convergence value of the sampling factor by performing a weighted non-linear least squares fitting of the data (see Fig.1b) using a function borrowed from electronics (this is a modified form of a limiting diode's equation) :

$$RMSD(s) = (s+c)\left(1 + \left|\frac{(s+c)}{a}\right|^b\right)^{-(1/b)}$$

where *(s)* is the sampling factor, *RMSD(s)* is the corresponding value of maximal RMSD, and *(a, b, c)* are the parameters whose values are to be determined through the non-linear least squares fitting procedure. The function chosen is especially useful because the value of the parameter *(a)* is directly related with the problem in hand and equals the expected value of max(RMSD) upon convergence. The use of the expression '*expected value*' in the previous sentence is important : if the trajectory is nowhere near convergence, then the [max(RMSD) vs. sampling] distribution will not reach convergence and the value of the parameter *(a)* will be higher than all observed –from the matrix– max(RMSD) values. This can –and is– being used as a criterion of insufficient sampling in the program we distribute. As is obvious from Fig.1b, the limiting diode equation fits exceedingly well both the cases of almost immediate convergence (black and cyan curves in Fig.1b), as well as the cases of slower convergence (red and green curves in Fig.1b). Once the value of the parameter *(a)* from the equation above is available [i.e. once we know the expected max(RMSD) upon convergence], an optimal value for the sampling factor can easily be determined by locating the smallest sampling factor which gives a max(RMSD) that is within 1σ (or higher) from this expected max(RMSD).

Once the proper sampling factor *($s_{optimal}$)* is known, the proposed method proceeds to completion smoothly : (1) Construct the *($s_{optimal}$)* different sub-matrices corresponding to the different choices of origin of the original matrix. (2) Use each of these sub-matrices with a hierarchical clustering method to calculate the corresponding dendrograms. (3) Use the dendrograms to calculate observed frequencies of 'molecular species' as a function of RMSD cutoff, (4) Apply Good-Turing statistics to determine the corresponding values of (probability of unobserved species vs. RMSD). (5) In the final step, average these *($s_{optimal}$)* different estimates of (probability of unobserved species vs. RMSD) and emit the final averages together with their estimated standard deviations.

Note that the method as described above is directly applicable to cases where instead of one long trajectory, several independent shorter runs were performed : As long as the initial configuration relaxes quickly (or, better still, is excluded from the calculations), the independent runs can be concatenated and the procedure described above applied without changes. Given the relatively coarse sampling of the trajectories needed by this method, the errors arising from the presence of time-dependent discontinuities at the connecting points should be negligible.

We should also note that although we have chosen to work with RMSD-based distance matrices, our method is also directly applicable using other formulations of distance. For example, it is possible to move away from Cartesian-space-based distances by using, for example, the principal components obtained from dihedral-PCA. In this approach, the elements of the distance matrix would be the Euclidean distances between the dPCA-derived principal components of the respective structures. The basic problem with such approaches is, of course, that the resulting units of distance will not have an immediate and easily visualized physical meaning.

Although the main product of our method is a graph of *($p_{unobserved}$ vs. RMSD)*, we have devised a much more economical (and easily quoted) probabilistic measure of convergence. This measure is an estimate of what is to be expected if the length of the simulation is doubled. In more detail, we calculate the answer to the following question : "What would be the value of the expected maximal RMSD (compared with the already recorded structures) if the simulation time is to be doubled ?". To say the same thing in other words, we want to estimate how much different would be the most different structure that we would observe if we doubled the simulation time. We will hereafter denote this estimate as 2T-RMSD. The value of 2T-RMSD is easily calculable from the ($p_{unobserved}$ vs. *RMSD*) distribution : what we want to calculate is the expected RMSD for the single structure that would differ the most (from those already observed) if we were to double the simulation time. Clearly, if we have observed a total of N samples (corresponding to a NxN matrix), the sought RMSD is the one corresponding to a value of $p_{unobserved}=1/N$ which can easily be computed directly from the RMSD matrix. To summarize, we believe that the value of 2T-RMSD is useful not only because it is easily quoted, but also because it successfully estimates the answer to a very pragmatic question : do the expected gains –for the sought analysis– worth the cost of doubling our computational effort ? (noting also how this is the inverse of the usual approach which is based on comparing the two halves of a trajectory to conclude –upon 'convergence'– that you could, after all, have used half of the computational effort already expended).

**2.2 Implementation details**

The method described above has been implemented in a fully automated program written using the *R programming language* of the *R* package for statistical computing.[6] See section §5 for program availability. Here we will only mention briefly the major *R* functions and packages used for implementing our method. The weighted non-linear least squares fitting of the limiting diode equation is performed with the function `nlsLM()` from the `minpack.lm` package[7]. The hierarchical clustering is performed with the function `hclust()` from the `fastcluster` package[8], using the complete linkage method[2] (other clustering methods and parameters have been examined and found to give essentially identical results). The clusters at various RMSD cutoffs are produced through the `cutree()` function. Given –for a specific RMSD cutoff– a number of clusters together with the number of their members (for each cluster), the probability of unobserved species is calculated[3,4] as $p_0=N_1/N$ where N1 is the number of clusters with only one member and N is the dimension of the respective matrix. All other calculations are performed with established functions provided by the *R* package.

The program we distribute differentiates between two distinct scenarios and can emit either of two different types of output. The first scenario concerns the possibility that the distribution of [max(RMSD) vs. sampling] as shown in Fig.1 has not reached convergence. This implies that for the given trajectory convergence can not be quantified. In this case no graph of *($p_{unobserved}$ vs. RMSD)* is produced, and the program emits a text message reading :

```
The maximal RMSDs  between the observed trajectory
structures  have  not converged. This implies that
the  length  of  the  given  trajectory  does  not
suffice for meaningfully  quantifying  convergence.
The only comment  that  can safely be made is that
upon  doubling  the  simulation  time  you  should
expect  to  observe  structures  that  differ from
those already observed by more than  approximately
                  XXX Angstrom.
```

where the value of 'XXX Ångström' is estimated from the value of 2T-RMSD (see previous section for definitions). The expression '*by more than*' in the program output shown above must be taken literally : if the [max(RMSD) vs. sampling] distribution has not converged, it is only possible to estimate a *lower* limit of the RMS deviation, not an upper. In the second scenario the value of *($s_{optimal}$)* can be determined, a graph (plus a table) of *($p_{unobserved}$ vs. RMSD)* is produced, and the program emits a text message containing the estimated value of 2T-RMSD :

```
The maximal RMSDs of the trajectory converged with
a sub-sampling factor of YY. The analysis suggests
that  the  most  different  structure  you  should
expect  to  observe  if  you  double  the simulation
time   will    differ   by    no    more   than
approximately  XXX +- ZZZ Angstrom   (RMSD)   from
            those already observed.
```

where the value of 'XXX ± ZZZ Ångström' above is estimated from the value of 2T-RMSD (see previous section for definitions).

# 3. Results

We have tested our method with an extensive set of molecular dynamics trajectories available to us. These trajectories (all in explicit solvent and with full PME-based electrostatics, performed with the program NAMD[9]) range from 5 μs-long folding simulations of the CLN025[10] and LytA-derived peptides[11], to 2 μs-long simulations of an α-Lactalbumin-derived peptide[12,13], to a 100 ns-long simulation of stable 4-α-helical bundle[14] and to a 50 ns-long simulation of a 1386-residue homohexameric protein[15]. The trajectories we tested cover the whole range from almost complete disorder (for example hepta-alanine[16]), to extreme stability (for example a variant of the Rop protein[17]). Clearly, the details of how the simulations were performed are irrelevant for the analysis reported here. What matters is the agreement (for a given trajectory) between this method and other methods of quantifying convergence, and, of course, its internal consistency and computational stability.

## 3.1 The method is internally consistent

The general appearance of the *($p_{unobserved}$ vs. RMSD)* distributions produced by this method is shown in Fig.2a in the form of two independent curves (black and magenta) corresponding to the results obtained from two different peptide folding simulations [CLN025 (lower black curve) and a LytA-derived peptide (upper magenta curve), see next section for comparison and discussion of the differences]. Both curves show the general characteristics that we would reasonably expect from the proposed method : At low values of RMSD, the probability values are high which signifies the fact that it is quite probable for the simulation to visit structures that although different in detail, are nevertheless quite similar to some of the structures already observed. As the RMSD increases, the probability monotonically decreases approaching asymptotically zero. The rate and the exact form with which the probability approaches zero depends on the properties of the trajectory (and the corresponding matrix) and not on the size of the protein. For example, results from the simulation of a very stable ~210 residue-long protein (a variant of the Rop protein) in its folded state showed that extremely small values of *$p_{unobserved}$* were reached at an RMSD of only ~1Å, much faster than the curves shown in Fig.2a which were derived from folding simulations of a 10-mer (CLN025) and a 14-mer (LytA) peptide.

In Fig.2b we examine what is possibly the most important criterion of internal consistency, namely, the expected dependence of the *($p_{unobserved}$ vs. RMSD)* distributions on the extend of sampling. For this calculation we used the same CLN025 trajectory as in Fig.2a, but we limited the calculation to (a) only 40 ns of simulation time (green curve in Fig.2b), (b) 400 ns of simulation time (red curve), and finally, (c) the whole 5 μs trajectory (black curve, identical with the curve shown in Fig.2a). In agreement with our expectations, 40 ns is too short a simulation time even for such a stable and fast folder as the CLN025 peptide. The result is a *($p_{unobserved}$ vs. RMSD)* distribution with high values of $p_{unobserved}$, showing clearly that the sampling is not sufficient and that significantly different structures must be expected if the simulation is to be continued [the value of 2T-RMSD (see section §2) was estimated to be 2.7±0.3Å]. In contrast with the 40 ns simulation (and in agreement with the known folding behaviour and folding timescale of the CLN025 peptide), the 400 and 5000 ns trajectories demonstrate significantly lower values of $p_{unobserved}$ and are quite similar. Their major differences are located at the high RMSD part of the diagram where the longer trajectory (black curve) gives noticeably lower values of $p_{unobserved}$ [the value of 2T-RMSD for the 5000 ns trajectory was 2.3±0.2 Å]. Having noted the rather small differences between the 400 and 5000 ns curves in Fig.2b, we shall not resist the temptation of noting just how clearly and quantitatively this method demonstrates how difficult it is to faithfully and accurately sample the folding landscape of proteins and peptides : increasing the computational effort by more than an order of magnitude hardly made a pronounced difference in the probabilistic estimates of $p_{unobserved}$. Seen in this light, the proposed method appears to be dependable and its estimates robust. Note, however, that due to the fact that CLN025 is a fast and stable folder (see section §3.2), the small differences between the 400 and 5000 ns curves are also a strong indication that all major conformations of the peptide have been sampled.

Given that this is a proper probabilistic method, we have the opportunity to perform an acid test on the validity of the derived *($p_{unobserved}$ vs. RMSD)* distributions. The principal idea is that you make a prediction using only the first half of a trajectory, and then compare the prediction with the actual results obtained from the second half (in other words, you estimate probabilities from the first half, and then compare them with the observed frequencies from the second half). The two lower curves in Fig.2c compare the predicted distribution (black curve) with the observed frequencies (orange curve) using the two halves of the CLN025 trajectory. The agreement between these two distributions is so outstandingly good that their comparison may create the wrong impression that this is the level of accuracy to be expected by this method irrespectively of the specific properties of the trajectory being examined. This is definitely not so. Because the CLN025 peptide is a fast and stable folder, 2.5 μs of simulation time was possibly adequate for meaningfully sampling its folding landscape, thus increasing the accuracy of the Good-Turing estimation. But for a trajectory that is far from being adequately sampled, the observed frequencies can –and, indeed, should be expected to– deviate very significantly from the prediction. To make sure that this point is not missed, in the same figure we also compare three graphs obtained from the LytA peptide (which is a very slow and erratic folder, see section §3.2). The magenta curve in Fig.2c is the $(p_{unobserved}$ vs. $RMSD)$ distribution as calculated by this method using the segment 0.0–1.3 μs of the LytA trajectory. The blue curve shows the actual frequencies observed in the 1.3–2.6 μs segment, and the gray curve are the actual frequencies observed in the 1.3–5.0 μs segment of the trajectory. Clearly, and in good agreement with common sense, predictions can only be as good as the data available. We will close this paragraph with what we consider to be an important side note : the excellent agreement between the observed and the expected (Good-Turing-derived) distributions for the case of a well-sampled trajectory, serves, we believe, as a direct and convincing validation of the choice to apply Good-Turing statistics to molecular dynamics simulations.

One last internal consistency check concerns the evolution and accuracy of the 2T-RMSD values as a function of simulation time. In the first calculation, and using again the CLN025 trajectory, we calculated the values of 2T-RMSD for simulation lengths of 80, 400, 800, 2400 and 5000 ns. Ignoring estimated standard errors, the corresponding values of 2T-RMSD were found to be 2.76 Å, 2.62 Å, 2.66 Å, 2.39 Å and 2.25 Å, in good agreement with the expected reduction of the 2T-RMSD values as sampling improves. Note, however, how the estimate increased (instead of monotonically decreasing) as we moved from the 400 to the 800 ns simulation. This is a healthy (and expected) behavior for a method whose predictions are based solely on the evidence available in hand, and arises from the incorporation of new information (previously unobserved peptide conformations) in the segment of the trajectory from 400 to 800 ns. A similar behavior can be seen in Fig.2b at values of RMSD of ~0.6 Å where the values of $p_{unobserved}$ for the 400 ns (red) curve are slightly lower than those obtained from the 5000 ns (black) curve. This behavior is discussed in more detail in section §4. In a second calculation, aiming to evaluate the accuracy of the 2T-RMSD values, we estimated the 2T-RMSD value using only the first half of the CLN025 trajectory, and then compared the prediction with the actual values observed in the second half of the trajectory. Using the first 2.5 μs of the trajectory, the value of 2T-RMSD was estimated to be 2.42±0.08 Å. The most different structure observed in the second half of the trajectory (2.5 – 5 μs) had an RMSD of 2.49 Å, in excellent agreement with the predicted value.

**3.2 The method is consistent with other established algorithms**

To demonstrate the consistency of this method with other established methods we shall discuss in more detail the specifics of the simulations that were used to prepare the two graphs shown in Fig.2a. The lower black curve was obtained from a 5 μs-long folding simulation of the CLN025 peptide[10]. CLN025 is known to be a fast and stable β-hairpin folder[18], and in agreement with these studies our trajectory contains more than ~50 folding/unfolding events (data not shown). On the other hand, the LytaA-derived peptide (upper magenta curve in Fig.2a) is a very slow and erratic folder, possibly due to the presence of significant energetic frustration in its folding landscape.[11] Indeed, our 5 μs trajectory of LytA contains only two relatively short folding/unfolding events, of which the second event is only partially correct (the alignment of the β-strands was offset by one residue, data not shown). Even at this level of analysis, Fig.2a clearly and correctly demonstrates the differences between the behavior of the two peptides : the $p_{unobserved}$ values for the CLN025 trajectory are throughout the RMSD range several times lower than those obtained from the LytA trajectory, signifying the better sampling (for the same amount of simulation time) of the much faster and stable folder. The 2T-RMSD values further underline and quantify the differences between the two trajectories, with CLN025 giving an estimate of 2.25Å, significantly lower than the value of 3.06Å obtained from LytA.

Not unexpectedly, the indications obtained by analyzing these same trajectories with other rigorous methods for quantifying convergence are in very good agreement with our results. To put this in numbers, we have calculated a rather strict (and unrelated with our method) measure of convergence, the dihedral-PCA-based eigenspace overlap[1] between the two halves of the trajectories (the dPCA analysis was performed with the programs carma[19] and grcarma[20]). For the CLN025 trajectory the overlaps between the one-, two-, three- and four-dimensional spaces defined by the respective principal components (of the two halves of the trajectory) were : 1D=0.96, 2D=0.70, 3D=0.90, and, 4D=0.96 indicating an excellent agreement between the information contained in the two halves, a fact which is usually taken to imply convergence. For the LytA trajectory the results from the same calculation were : 1D=0.66, 2D=0.45, 3D=0.36, and, 4D=0.59, demonstrating again the rather incomplete sampling for the slow folder. We will note here how much more complete and satisfying is the probabilistic treatment of Fig.2a compared with the simple enumeration of eigenspace overlaps that beg for additional (possibly arbitrary) interpretations (for example, is the overlap of 0.59 between the two halves of the trajectory an indication that the full trajectory is adequately sampled ? And what "adequately" means, and how is it to be quantified, etc.). Since we are on this subject, and in the form of a side note, we show below how badly a popular measure of convergence, the cosine content, performs for this problem. The values of the cosine content for the first four dPCA-derived principal components of the LytA trajectory were 0.0333, 0.0040, 0.0291 and 0.0032, falsely indicating –according to popular interpretations– that convergence has been achieved. To make matters worse, the results from the cosine content analysis of LytA are practically indistinguishable from the values obtained from the CLN025 trajectory which were 0.0049, 0.0279, 0.0002 and 0.0001 respectively.

We close this section with an example which we believe demonstrates the dependability of our method. Fig.3 shows a series of graphs of the type [RMSD from the starting (crystal) structure vs. Simulation time] for the simulation of a large protein in the folded state[15]. Such graphs are very common in the literature and their appearance (mainly flatness of the distributions) is being advertised –and used– as an indication of convergence of sampling. This approach is clearly questionable since the RMSD from a given reference structure does not contain information about the actual extend of sampling of the configurational space available to the macromolecule being simulated. The example we selected brings this forward. The red curve in the upper panel of Fig.3 shows the evolution of the (RMSD vs. crystal structure) for all $C_\alpha$ atoms of BcZBP, a large 1386 residue-long homohexameric protein. The lower black curve in this same panel shows again a graph of the type (RMSD vs. crystal structure), but this time after excluding from the calculation ~30 residues (from each monomer) belonging to hyper-mobile surface-exposed loops. The lower panel in Fig.3 shows results from the same set of calculations, but this time using only one monomer (monomer C), instead of the whole hexamer. Risking a prediction, we believe that these graphs would be accepted as implying that (a) the whole hexamer and the whole monomer show no sign of convergence, (b) that the hexamer without the flexible loops is possibly approaching convergence, and finally, (c) the dynamics of the monomer without the flexible loops have converged. The results from our method come as something of an anticlimax : for none of these trajectories the method even reached the stage of estimating *($p_{unobserved}$ vs. RMSD)*. All trajectories, including the monomeric-no-loops trajectory (black curve in the lower panel of Fig.3), failed to even pass the criterion of convergence of the *(max(RMSD) vs. sampling)* distribution. Clearly, a method that justifiably defies common misconceptions is probably a dependable (though possibly unpopular) tool.

## 3.3 The method is robust and insensitive to sampling choices

The computer memory requirements for this method are so demanding that only matrices of the order of few thousand can be examined, with matrices of approximately 20,000×20,000 being the upper limit for the current generation of workstations, and matrices of approximately 10,000×10,000 the norm. This posses the question of how sensitive is the proposed method to the arbitrary choices that can be made during the sampling of the trajectory. There are two aspects of the problem. The first is the sensitivity of the results on the 'step size' with which the trajectory was sampled. The second aspect is the dependence (for a given 'step size') of the results on the choice of the frame that was selected to be the first frame of the matrix (i.e. the frame offset). Here we examine both of these questions using a 100 ns simulation of a 216-residue-long 4-α-helical bundle protein (a homotetrameric variant of the Rop protein[14]). The complete trajectory contained 250,000 frames and for the calculations reported here we only used the protein's $C_\alpha$ atoms.

The first set of calculations concerns the sensitivity of the *($p_{unobserved}$ vs. RMSD)* distributions on the step size used for sampling the original trajectory. To establish how robust (or otherwise) the estimates are, we have chosen to also use some unreasonably large sampling factors. In this spirit we tested all step sizes ranging from 20 frames per step (giving a 12500×12500 matrix), all the way to a step size of 100 (giving a matrix of only 2,500×2,500), with an interval of 10 frames for the values in between. Fig.4a shows a superposition of the *($p_{unobserved}$ vs. RMSD)* distributions obtained from these nine step sizes (note that the scaling of the horizontal axis has been changed to magnify the differences). In general the agreement between the distributions obtained from the various step sizes is excellent. What is even more reassuring, however, is the behavior of the method as the sampling of the trajectory becomes artificially coarse : The curves that appear to deviate significantly from the bulk (gray and cyan curves in Fig.4a) are those that correspond to these artificially coarse samplings, and for these distributions the $p_{unobserved}$ probabilities are noticeably *overestimated*. The implication is clear : the proposed method is safe and robust –in the sense that the derived probabilities are bounded from below– even in the cases where the sampling of the original trajectory is unreasonable.

The second set of calculations examines (for a fixed value of step size) the sensitivity of the *($p_{unobserved}$ vs. RMSD)* distributions on the offset used to select the first frame that will enter the RMSD matrix. For this calculation we used a fixed step size of 70 (giving a matrix of 3571×3571), a fixed value for the sampling factor of 2, and tested all combinations of first frame ranging from 1 to 66 with a step of 5, giving a total of 14 combinations. Because in the limit of an infinite trajectory the choice of the first frame would have no effect on the derived curve, the variability that is observed in these graphs is a fair representation of the statistical noise present in the procedure. The resulting graphs are shown in Fig.4b, and as expected, the effect of frame offset is negligible for a constant value of the sampling factor. It should be noted, however, that due to the presence of statistical noise, the algorithm that selects the value of the sampling factor *(s)* as described in section §2.1, may select a slightly different sampling factor depending on the value of the frame offset which, in turn, would lead to a slightly different *($p_{unobserved}$ vs. RMSD)* curve. This, again, is safe in the sense that the probabilities are bounded from below, and any deviations always lead to *overestimating* the $p_{unobserved}$ probabilities.

# 4. Discussion

We have shown that quantifying convergence of molecular dynamics trajectories in probabilistic terms not only is feasible, but also that its application resulted in the development of a method that appears to be consistent, robust and dependable. We have also shown that the expected major deficiency of the algorithm, that is, the very limited number of structures that can be used for the analysis, appears to be a non-existent problem in the sense that for the time scales of present day simulations we usually have to sub-sample even further the original RMSD matrices.

What would appear to be the most basic problem with the proposed method is the fundamental idea that clusters of structures (at a given RMSD level) can be treated as multiple observations of the same 'object' taken from the discrete distribution. The excellent agreement between the observed and the expected (Good-Turing-derived) distributions shown in Fig.2c for a well-sampled trajectory clearly indicates that this is a valid approximation, but it could still be argued that what is needed in this case is a formalism similar to Good-Turing, but for the continuous distribution case –which to our knowledge is not available. On the other hand, even if such a continuous-case treatment did exist, we would then face the opposite problem, which is that the recording of structures in our trajectories is indeed discrete (and not continuous). Additionally, this view of macromolecular conformations as (possibly numerous) groups of distinct clusters maybe relevant biologically, and is (at least metaphorically) consistent with the idea of the existence of local roughness in the energy landscape of proteins.

Turning our attention to practical aspects of the day-to-day application of the method, we should probably start from a semi-philosophical cliché : The probabilistic treatment described above, although consistent and satisfying, is not a panacea. The probabilities calculated by this method are based solely on the evidence in hand, and there is no way for this method (or any other method) to "guess", for example, that if a given simulation was continued for, say, another 50 ns we would then observe several new stable conformations that would force us to revise (upwards) the $p_{unobserved}$ estimates. Similarly, there is no way for the method to "guess", for example, that a stable conformation that lasted for more than 90% of the time of a 20 ns trajectory (and led to very low values of $p_{unobserved}$) would turn-out to be statistically insignificant if the simulation were to be extended to 2000 ns. Having said that, in none of our tests with tens of different trajectories have we observed a systematic flaw –in the form of unjustifiable predictions– made by this method.

The second most important thing that we should like to note concerning the practical application of the method has more to do with the human factor, and almost nothing with the method *per se*. What we refer to is, of course, that it is very tempting to prefer to look at a graph like the one shown in Fig.3 (black curve in the lower panel) and to conclude that "convergence has been achieved", instead of applying a probabilistic method that returns a message in the spirit of "insufficient sampling, double the simulation time and try again". This is exacerbated by the apparent 'honesty' of the method, whose predictions appear some times to be disheartening with respect to the amount of computational effort they imply that is required to improve sampling (compare, for example, the red and black curves in Fig.2b corresponding to 400 vs. 5000 ns of simulation time). Clearly, and as with any new tool, only accumulated practical experience with the method will show what probability level is to be considered significant for a given problem.

We close this section with an aphorism. We believe that what this method clearly demonstrated is that there is no such thing as a positive declaration of 'convergence' or 'sufficient sampling'. In full agreement with common sense, our method showed that all that is happening as simulation time increases is that the probability of encountering new –thus far unobserved– conformations asymptotically decreases (and that, thus, the trust we place upon the conclusions drawn from the trajectory must increase). Not unexpectedly, treating an inherently probabilistic problem probabilistically leads to predictions that are in excellent agreement with common sense.

## 5. Program availability

A free and open source program, published under a Simplified BSD License, which implements the method described in this communication is immediately available for download *via* the github repository at https://github.com/pkoukos/GoodTuringMD

# 6. References

1. Grossfield, A.; Zuckerman, D. M. Quantifying Uncertainty and Sampling Quality in Biomolecular Simulations. *Annu. Rep. Comput. Chem.* **2009,** *5*, 23-48.

2. Shao, J.; Tanner, S. W.; Thompson, N.; Cheatham, T. E., III. Clustering Molecular Dynamics Trajectories: 1. Characterizing the Performance of Different Clustering Algorithms. *J. Chem. Theory Comput.* **2007,** *3*, 2312–2334.

3. Good, I. J. The population frequencies of species and the estimation of population parameters. *Biometrika* **1953,** *40*, 237-264.

4. Gale, W. A.; Sampson, G. Good-Turing Frequency Estimation Without Tears. *Journal of Quantitative Linguistics* **1995,** *2,* 217-237.

5. Flyvbjerg, H.; Petersen, H. G. Error estimates on averages of correlated data. *J. Chem. Phys.* **1989,** *91*, 461-467.

6. R Core Team. *R: A language and environment for statistical computing,* 3.0.0; R Foundation for Statistical Computing: Vienna, Austria, 2013.

7. Elzhov, T. V.; Mullen, K. M.; Spiess, A. N.; Bolker, B. *minpack.lm: R interface to the Levenberg-Marquardt nonlinear least-squares algorithm found in MINPACK, plus support for bounds*, 1.1-7; R Foundation for Statistical Computing: Vienna, Austria, 2013.

8. Mullner, D. *fastcluster: Fast hierarchical clustering routines for R and Python*, 1.1.9; R Foundation for Statistical Computing: Vienna, Austria, 2013.

9. Phillips J. C.; Braun, R.; Wang, W.; Gumbart, J.; Tajkhorshid, E.; Villa, E.; Chipot, C.; Skeel, R. D.; Kale, L.; Schulten, K. Scalable molecular dynamics with NAMD. *J. Comput. Chem.* **2005,** *26*, 1781-1802.

10. Hatfield, M. P. D.; Murphy, R. F.; Lovas, S. Molecular dynamics analysis of the conformations of a β-hairpin miniprotein. J. Phys. Chem. B **2010,** *114*, 3028-3037.

11. Patmanidis, I.; Glykos, N. M. As good as it gets? Folding molecular dynamics simulations of the LytA choline-binding peptide result to an exceptionally accurate model of the peptide structure. *J. Mol. Graph. Model.* **2013,** *41*, 68-71.

12. Patapati, K. K.; Glykos, N. M. Order through Disorder: Hyper-Mobile C-Terminal Residues Stabilize the Folded State of a Helical Peptide. A Molecular Dynamics Study. PLoS One **2010,** *5*, e15290.

13. Patapati, K. K.; Glykos, N. M. Three Force Fields' Views of the 310 Helix. *Biophys. J.* **2011,** *101*, 1766-1771.

14. Glykos, N. M.; Papanikolau, Y.; Vlassi, M.; Kotsifaki, D.; Cesareni G.; Kokkinidis, M. Loopless Rop: Structure and Dynamics of an Engineered Homotetrameric Variant of the Repressor of Primer Protein. *Biochemistry* **2006,** *45*, 10905-10919.

15. Fadouloglou, V. E.; Stavrakoudis, S.; Bouriotis, V.; Kokkinidis, M.; & Glykos, N. M. Molecular Dynamics Simulations of BcZBP, A Deacetylase from Bacillus cereus: Active Site Loops Determine Substrate Accessibility and Specificity. *J. Chem. Theory Comput*. **2009,** *5*, 3299-3311.

16. Georgoulia, P. S.; Glykos, N. M. Using J-coupling constants for force field validation: Application to hepta-alanine. *J. Phys. Chem. B* **2011,** *115*, 15221–15227.

17. Glykos, N. M. On the application of molecular-dynamics simulations to validate thermal parameters and to optimize TLS-group selection for macromolecular refinement. *Acta Crystallogr.* **2007,** *D63*, 705-713.

18. Zhao, G. J.; Cheng, C. L. Molecular dynamics simulation exploration of unfolding and refolding of a ten-amino acid miniprotein. *Amino Acids* **2012,** *43*, 557-565.

19. Glykos, N. M. Carma: a molecular dynamics analysis program. *J. Comput. Chem.* **2006,** *27*, 1765-1768.

20. Koukos, P. I.; Glykos, N. M. grcarma: A Fully Automated Task-Oriented Interface for the Analysis of Molecular Dynamics Trajectories. *J. Comput. Chem.* **2013**. DOI: 10.1002/jcc.23381.

# Figure Captions

**Figure 1**

**Determination of the sampling factor :** The two curves in the upper panel compare the behavior of the [max(RMSD) vs. sampling factor] distribution (see text for details) for a reasonably sampled trajectory (upper black curve), and for a trajectory sampled so finely that successive structures are highly correlated (lower red curve). For the upper curve, the original RMSD matrix would have to be sub-sampled with a sampling factor of 2, whereas for the lower (red) curve a much higher sampling factor of ~30 would be necessary. Panel (b) illustrates how well the limiting diode equation (see text for details) can fit the data. The underlying data are the same with panel (a) and the two curves (cyan and green) are the corresponding least squares fits.

**Figure 2**

**General form of the results, dependence on the extend of sampling and internal consistency :** Panel (a) shows the general form of the distributions *($p_{unobserved}$ vs. RMSD)* obtained by this method using two independent 5 μs-long folding simulations of the CLN025 peptide (black curve) and the LytA-derived peptide (magenta curve, see text for details). Panel (b) shows how the estimated probabilities change as the length of the simulation of the CLN025 peptide increases from 40 ns (green curve), to 400 ns (red curve), to 5000 ns (black curve). Finally, in panel (c) we compare the expected and observed forms of the *($p_{unobserved}$ vs. RMSD)* distribution as obtained from two trajectories. The two lower (black and orange) curves are based on the CLN025 trajectory and compare the expected distribution as calculated using only the first half of the trajectory (black curve), with the actual frequencies observed in the the second half (orange curve). The upper set of (three) curves are based on the LytA trajectory (see text for details) and compare : (a) the expected distribution (colored magenta) as calculated from this method using only the first 0–1.3 μs part of the trajectory, (b) the distribution actually observed in the 1.3–2.6 μs part of the trajectory (blue curve), and, (c) the distribution actually observed in the 1.3–5.0 μs part of the trajectory (gray curve).

**Figure 3**

**Comparison with other methods :** The two panels show the evolution of (RMSD vs. crystal structure) as a function of simulation time for either the whole hexamer of the BcZBP protein (upper panel), or only one monomer (monomer C, lower panel), see text for details. In each panel, the upper (red) curves were calculated using all $C_\alpha$ atoms, the lower (black) curves after excluding residues belonging to hyper-mobile surface-exposed loops. For all four cases shown, the method described in this communication returned the 'insufficient sampling' message as described in section §2.2.

**Figure 4**

**Insensitivity to trajectory sampling choices :** Panel (a) shows the dependence of the *($p_{unobserved}$ vs. RMSD)* distributions on the step size used for sampling the original 100 ns-long trajectory of a stable 4-α-helical bundle protein. The superposition of nine curves (corresponding to step sizes ranging from every 20 to every 100 frames, with an interval of 10 frames) are shown. The gray and cyan curves (to the left) correspond to the (unreasonably) large step sizes. Panel (b) shows a superposition of *($p_{unobserved}$ vs. RMSD)* curves which were obtained by keeping the step size constant, and varying the trajectory frame which was taken to be the first (i.e. the frame offset).

# Figure 1

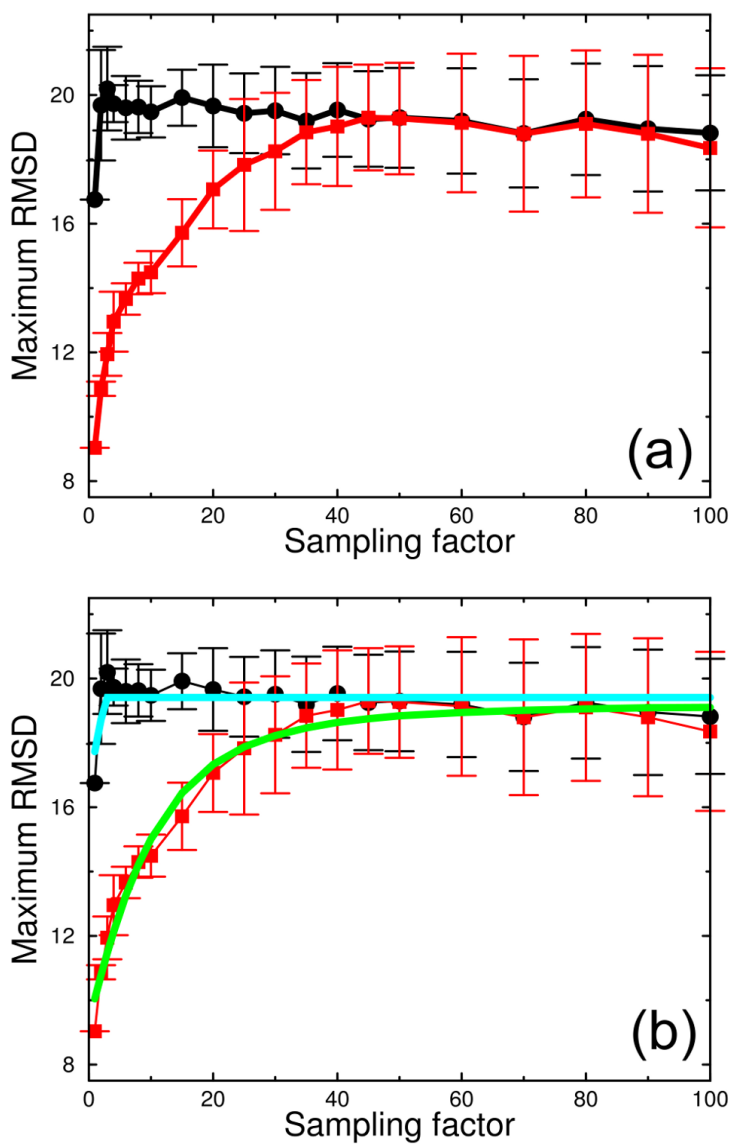

**Determination of the sampling factor :** The two curves in the upper panel compare the behavior of the [max(RMSD) vs. sampling factor] distribution (see text for details) for a reasonably sampled trajectory (upper black curve), and for a trajectory sampled so finely that successive structures are highly correlated (lower red curve). For the upper curve, the original RMSD matrix would have to be sub-sampled with a sampling factor of 2, whereas for the lower (red) curve a much higher sampling factor of ~30 would be necessary. Panel (b) illustrates how well the limiting diode equation (see text for details) can fit the data. The underlying data are the same with panel (a) and the two curves (cyan and green) are the corresponding least squares fits.

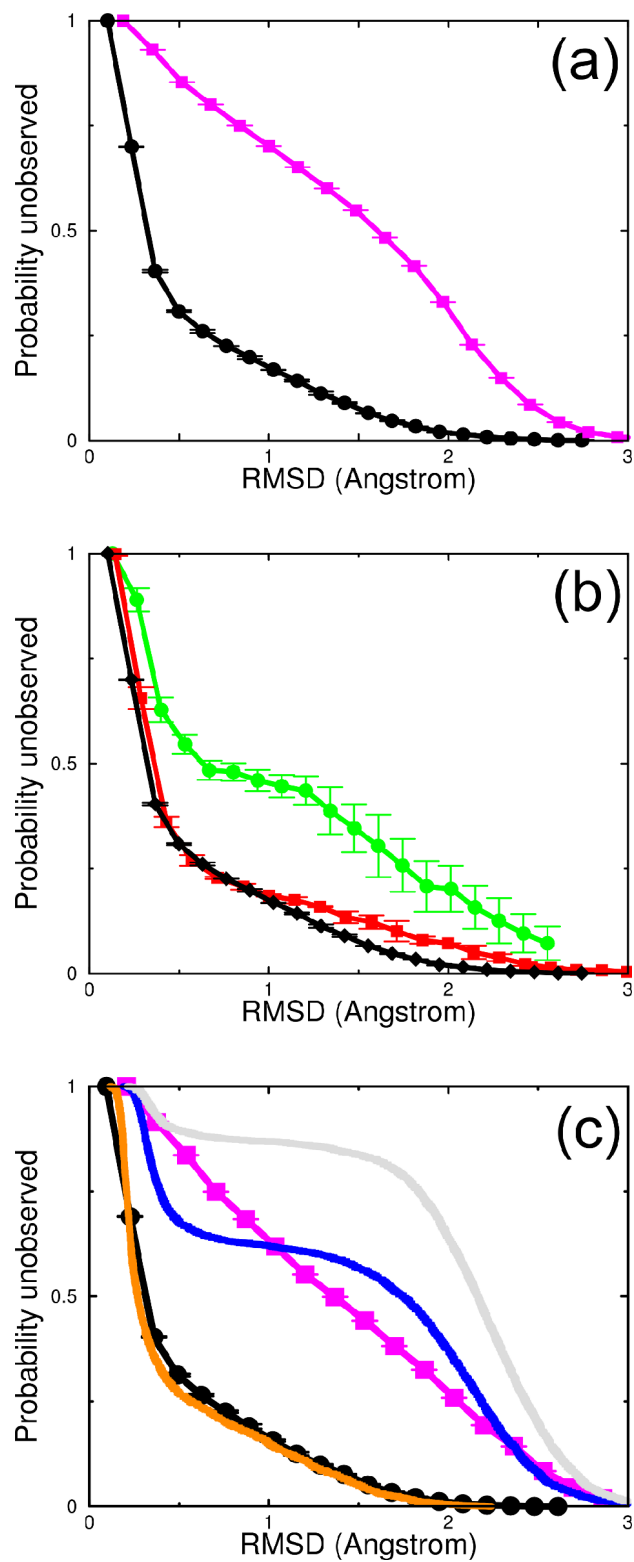

**General form of the results, dependence on the extend of sampling and internal consistency :**
Panel (a) shows the general form of the distributions *($p_{unobserved}$ vs. RMSD)* obtained by this method using two independent 5 μs-long folding simulations of the CLN025 peptide (black curve) and the LytA-derived peptide (magenta curve, see text for details). Panel (b) shows how the estimated probabilities change as the length of the simulation of the CLN025 peptide increases from 40 ns (green curve), to 400 ns (red curve), to 5000 ns (black curve). Finally, in panel (c) we compare the expected and observed forms of the *($p_{unobserved}$ vs. RMSD)* distribution as obtained from two trajectories. The two lower (black and orange) curves are based on the CLN025 trajectory and compare the expected distribution as calculated using only the first half of the trajectory (black curve), with the actual frequencies observed in the the second half (orange curve). The upper set of (three) curves are based on the LytA trajectory (see text for details) and compare : (a) the expected distribution (colored magenta) as calculated from this method using only the first 0–1.3 μs part of the trajectory, (b) the distribution actually observed in the 1.3–2.6 μs part of the trajectory (blue curve), and, (c) the distribution actually observed in the 1.3–5.0 μs part of the trajectory (gray curve).

# Figure 3

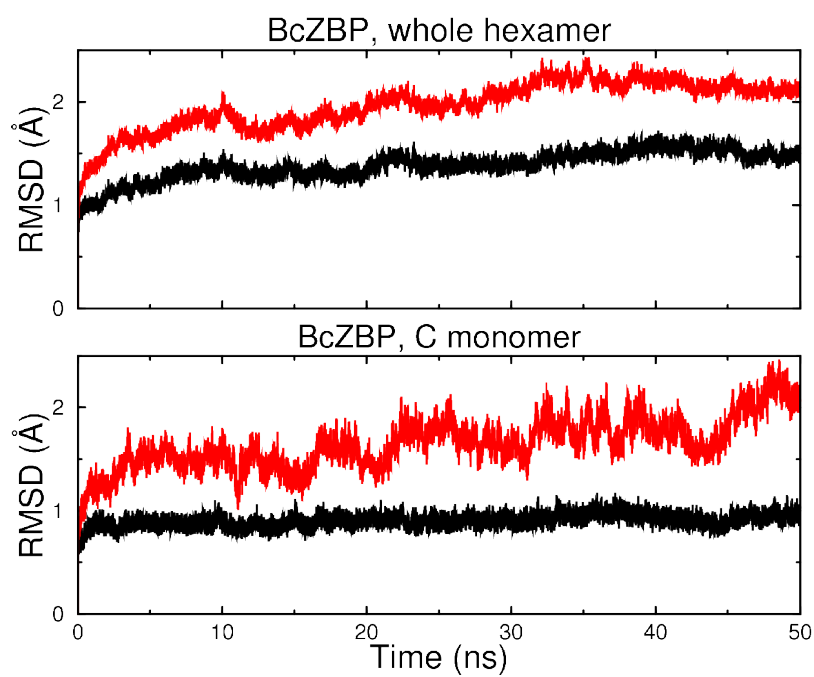

**Comparison with other methods :** The two panels show the evolution of (RMSD vs. crystal structure) as a function of simulation time for either the whole hexamer of the BcZBP protein (upper panel), or only one monomer (monomer C, lower panel), see text for details. In each panel, the upper (red) curves were calculated using all $C_\alpha$ atoms, the lower (black) curves after excluding residues belonging to hyper-mobile surface-exposed loops. For all four cases shown, the method described in this communication returned the 'insufficient sampling' message as described in section §2.2.

**Figure 4**

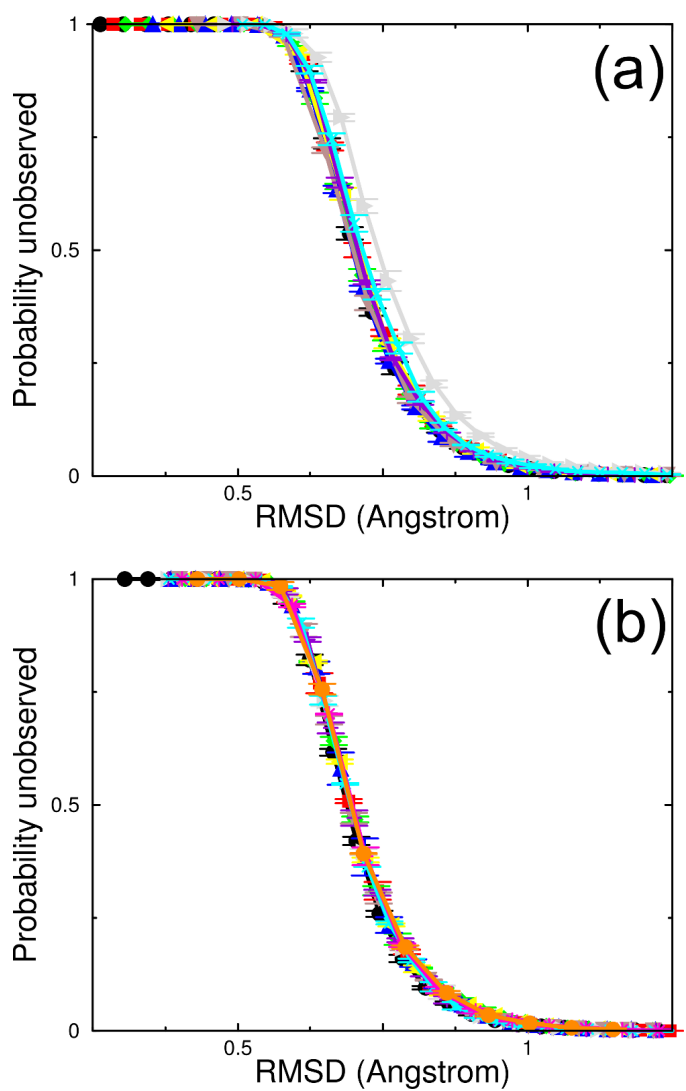

**Insensitivity to trajectory sampling choices :** Panel (a) shows the dependence of the *($p_{unobserved}$ vs. RMSD)* distributions on the step size used for sampling the original 100 ns-long trajectory of a stable 4-α-helical bundle protein. The superposition of nine curves (corresponding to step sizes ranging from every 20 to every 100 frames, with an interval of 10 frames) are shown. The gray and cyan curves (to the left) correspond to the (unreasonably) large step sizes. Panel (b) shows a superposition of *($p_{unobserved}$ vs. RMSD)* curves which were obtained by keeping the step size constant, and varying the trajectory frame which was taken to be the first (i.e. the frame offset).